\documentclass[prd,preprintnumbers,superscriptaddress,nofootinbib,showpacs,twocolumn]{revtex4-2}
\usepackage{pstricks}
\usepackage{color}
\usepackage{amssymb,amsmath,bm,bbm}
\usepackage{epsf,epsfig}
\usepackage{afterpage}
\usepackage{longtable}
\usepackage{latexsym,mathrsfs,dsfont}
\usepackage{graphics}
\usepackage{url}
\usepackage{paralist}
\usepackage{bbold}
\usepackage{appendix}
\usepackage{tikz}
\usepackage[colorlinks=true, urlcolor=blue]{hyperref}

\newcommand{\be}{\begin{equation}}
\newcommand{\ee}{\end{equation}}
\newcommand{\bi}{\begin{itemize}}
\newcommand{\ei}{\end{itemize}}
\newcommand{\ba}{\begin{array}}
\newcommand{\ea}{\end{array}}
\newcommand{\bea}{\begin{eqnarray}}
\newcommand{\eea}{\end{eqnarray}}
\newcommand{\bec}{\begin{center}}
\newcommand{\eec}{\end{center}}

\newcommand{\nn}{\nonumber}

\def\@seccntformat#1{\@ifundefined{#1@cntformat}%
   {\csname the#1\endcsname\quad}
   {\csname #1@cntformat\endcsname}
}

\begin{document}

\preprint{BARI-TH/760-24}

\title{   
Constraining  $\nu$SMEFT coefficients: the case of the extra $\text{U}(1)^\prime$ }
\author{Pietro~Colangelo}
\email[Electronic address:]{pietro.colangelo@ba.infn.it} 
\affiliation{Istituto Nazionale di Fisica Nucleare, Sezione di Bari, via Orabona 4, 70126 Bari, Italy}
\author{Fulvia~De~Fazio}
\email[Electronic address:]{fulvia.defazio@ba.infn.it} 
\affiliation{Istituto Nazionale di Fisica Nucleare, Sezione di Bari, via Orabona 4, 70126 Bari, Italy}
\author{Francesco~Loparco}
\email[Electronic address:]{francesco.loparco1@ba.infn.it} 
\affiliation{Istituto Nazionale di Fisica Nucleare, Sezione di Bari, via Orabona 4, 70126 Bari, Italy}
\author{Nicola~Losacco}
\email[Electronic address:]{nicola.losacco@ba.infn.it} 
\affiliation{Istituto Nazionale di Fisica Nucleare, Sezione di Bari, via Orabona 4, 70126 Bari, Italy}
\affiliation{Dipartimento Interateneo di Fisica "Michelangelo Merlin", Universit\`a degli Studi di Bari, via Orabona 4, 70126 Bari, Italy}

\begin{abstract}
\noindent 
We study the constraints on  low-energy coefficients of the $\nu$SMEFT generalization of the Standard Model effective theory in the simple case of a $\text{U}(1)^\prime$ enlargement of the Standard Model gauge group. 
In particular, we analyse the constraints imposed by the requirement that the extended theory remains free of gauge anomalies. We present the cases of explicit realisations, showing the obtained correlations among the coefficients of $d=6$ operators.

\end{abstract}

\thispagestyle{empty}


\maketitle
\section{Introduction}
 The search for physics beyond the Standard Model (SM) is justified by several motivations.
There are conceptual issues and cosmological observations suggesting the existence of  a more fundamental theory beyond SM.
Tensions between SM predictions and experimental results,  in particular in the flavour sector, reinforce such a widespread conviction.
However, direct searches at colliders have not produced evidence of  new particles and/or mediators of new interactions, yet, hence the alternative  way to gain evidence of physics beyond the Standard Model (BSM)  is investigating virtual effects of  possible new heavy degrees of freedom, as done in
flavour physics  \cite{DeFazio:2023lmy}. 

In this framework, two approaches can be followed towards BSM.
The first one consists in formulating a specific extended theory and  deriving predictions to be contrasted with experiment for a  validation or  a discrimination with respect to different new physics  (NP) scenarios. 
The second approach consists in extending the SM at the electroweak (EW) scale in the most general way compatible with the SM gauge symmetry,  investigating the constraints imposed by the experiments on the resulting generalization.

A  remarkable example of the second approach is the Standard Model effective field theory (SMEFT) \cite{Buchmuller:1985jz,Grzadkowski:2010es,Brivio:2017vri,Isidori:2023pyp},  widely used  in the quest for BSM physics.
The SM is considered as an effective field theory describing physics at and below the EW scale.
At higher scales a new gauge theory (the UV completion) should exist, with a gauge group extending the SM one and  undergoing spontaneous symmetry  breaking (SSB) to it. 
If  $\Lambda$ is the NP scale, at the EW scale the SMEFT Lagrangian consists of an expansion in the  parameter $ 1/\Lambda$.
The first term of the expansion is the SM Lagrangian density  containing operators of canonical dimension  up to $d=4$. 
Subsequent terms are suppressed by  powers of  $1/\Lambda$  and comprise operators of increasing dimension:
\be
{\cal L}_{\rm SMEFT}={\cal L}_{\rm SM}^{(4)}+{\cal L}^{(5)}+{\cal L}^{(6)}+ \dots \;. \label{LSMEFT}
\ee
The apex $(d)$ indicates the canonical dimension of the operators entering in each term ${\cal L}^{(d)}$   written as 
\be
{\cal L}^{(d)}=\sum_i \frac{C_i}{\Lambda^{d-4} } \, {\cal O}_i^{(d)} \;, \label{Ld}
\ee 
with  dimensionless Wilson coefficients $C_i$.
The operators are constructed in terms of the SM fields and satisfy the SM gauge symmetry. SM accidental symmetries are allowed to be violated:
for example, baryon and lepton number violating operators are included in \eqref{Ld}, namely odd-dimension operators violating $B$ and/or $L$ conservation \cite{Helset:2019eyc}. 
The operators contain no reference to the field content of the UV theory.
However, their coefficients depend of the details of such a theory, i.e. the couplings and masses of the new particles that, supposed to be  $M \simeq{\cal O}(\Lambda)$, are  integrated out  in the effective field theory (EFT) Lagrangian at the EW scale.
A few assumptions  concern the UV theory. It  should contain only particles with spin $J \le 1$; new vector fields could be either gauge fields (massless before  SSB in the UV theory) or massive Proca fields; new fermions can be introduced provided that they are vector-like with respect to the SM gauge group,  to maintain the SM  free of  gauge anomalies. Even though in the construction of the SMEFT operators the latter requirement is taken into account, in more general frameworks it can be relaxed, provided that together with new fermions (not necessarily vector-like), other contributions are added that maintain the SM gauge group anomaly free \cite{Dekens:2019ept}.

One can use the construction in two ways.
Choosing the UV completion, the Wilson coefficients of the SMEFT operators can be  determined through   matching and running procedure \cite{Jenkins:2013zja,Jenkins:2013wua,Alonso:2013hga}.
On the other hand, without assumptions on the UV completion, the coefficients are treated as  parameters. 
These two steps are complementary to each other.  Having gained model independent information on the  coefficients in the effective theory, it is possible to contrast them with the features required in a specific scenario in order to validate or discard it.

The phenomenological  evidence that neutrinos  have nonvanishing mass induces to  consider the $\nu$SMEFT  extension of SMEFT,   which comprises  three right-handed sterile neutrino fields in the  sub-TeV mass range \cite{delAguila:2008ir,Aparici:2009fh,Bhattacharya:2015vja,Liao:2016qyd,Alcaide:2019pnf,Chala:2020pbn,Datta:2020ocb,Datta:2021akg}. 
The inclusion  does not invalidate the requirement that the SM  gauge group is  free of gauge anomalies. 
 In the extension, ${\cal L}^{(5)}$ consists of three operators, while only the Weinberg operator appears at this order  in the absence of  $\nu_R$  \cite{Weinberg:1979sa}.
The choice of the $d=6$  operators  is not unique, and different {\it bases} have been proposed, i.e. complete sets of independent,  non redundant operators.\!\!\!\!
\footnote{Sources of redundancies  are, e.g.,  operators  obtained one from the other after integration by parts and discarding a total derivative; operators that can be discarded using equations of motion; equivalent operators  upon Fiertz transformations (in the case of four-fermion operators).} 
A popular basis is the Warsaw one \cite{Grzadkowski:2010es}. 
In each basis the operators are collected in classes according to their field content. 

In our study we focus on the UV completion represented by the simplest extension of the SM gauge group  comprising a new $\text{U}(1)^\prime$ gauge group,  featured by the gauge coupling $g_Z$ \cite{Leike:1998wr,Langacker:2000ju,Appelquist:2002mw,Rizzo:2006nw}.
$Z^\prime$ is the corresponding gauge field, and the $z$-hypercharge is the quantum number associated to the new symmetry.
Many NP models  introduce such a  mediator  with specific $z$-hypercharge assignments. 
Experimental searches for $Z^\prime$ rely on the assumptions for the  hypercharges, and   produce  exclusion plots in the plane of  the $Z^\prime$ production cross section versus   $M_{Z^\prime}$. \!\!\!\!\!\!
\footnote{See, e.g.,  the review: B.A. Dobrescu and S. Willocq, "$Z^\prime$-boson searches", in \cite{Navas:PDG}. }
The NP scale can be identified with  $M_{Z^\prime}$ acquired after spontaneous  breaking  of the new symmetry. 
We do not need to specify how such SSB occurs, we only assume that it happens at a  much higher scale than the SM Higgs vacuum expectation value. We  neglect the mixing with  other neutral gauge bosons.\!\!\!
\footnote{Mixing at tree-level vanishes in models where the SM Higgs is assumed to be singlet under $\text{U}(1)^\prime$.}

In the chosen extension we work out  the coefficients of the $\nu$SMEFT operators of dimension up to $d=6$, aiming at   the relations  among them.\!\!\!\!
 \footnote{In the same framework, relations among the coefficients of $d=6$ and  $d=8$ operators have been worked out in \cite{Dawson:2024ozw}.}
While the gauge structure of the theory already imposes nontrivial relations among  various coefficients,  further  relations can be established  requiring that the extended gauge group is anomaly free. 
We obtain results  holding for a generic $\text{U}(1)^\prime$ extension. 
 We also consider specific cases: universal $Z^\prime$ couplings to the three generations or  only to the third generation; $Z^\prime$ only coupled  to left- or right-handed fermions;  lepto-  or hadrophobic $Z^\prime$;  the $z$-hypercharge assignment of the ABCD model \cite{Aebischer:2019blw}. 
 In all cases, we find that the number of independent coefficients is reduced and  remarkable correlations can be established among them, which are peculiar of each extension. 
 The experimental test of such correlations would shed light on the particular completion,  providing the widest information using measurements.

The plan of the paper is as follows.  
After  Sec.~\ref{setup} with the  notations,   
in Sec.~\ref{smeft} we list the $\nu$SMEFT operators generated at the EW scale when the UV theory contains the new gauge boson $Z^\prime$. 
The impact of the new gauge boson on the SMEFT Lagrangian density is considered in Sec.~\ref{relations}, with the list of the operators  obtained when the $Z^\prime$ field is integrated out, the expressions of their Wilson coefficients and the relations  due to the gauge structure of the  extension. 
In Sec.~\ref{anomalies} we consider the relations that the fermion $z$-hypercharges must satisfy to fulfil the requirement of gauge anomaly cancellation in the SM gauge group  extension,  and how such relations can be translated into analogous ones among the SMEFT coefficients.  
 We than discuss the results for the selected $z$-hypercharge assignments. 
The last section  comprises the conclusions.

\section{Notations,    $Z^\prime$ couplings to fermions and to the Higgs field }\label{setup}
 The most general renormalizable UV Lagrangian terms involving the  gauge boson $Z^\prime$ of a new $U(1)^\prime$ group can be written as
\be
{\cal L}^{Z^\prime}={\cal L}^{Z^\prime}_{\rm free}+{\cal L}^{Z^\prime}_{\rm int, fermions}+{\cal L}^{Z^\prime}_\varphi
\,\,. \label{LZp} 
\ee
The first term in \eqref{LZp} reads
\be
{\cal L}^{Z^\prime}_{\rm free}=-\frac{1}{4}Z^\prime_{\mu \nu} Z^{\prime \mu \nu} +\frac{1}{2}M^2_{Z^\prime}
Z^\prime_\mu  Z^{\prime \mu } \,\, .
\ee
$Z_\mu^\prime$ is the gauge boson field and $Z^\prime_{\mu \nu}=\partial_\mu Z^\prime_\nu -\partial_\nu Z^\prime_\mu$ is the  field strength tensor.

The second term in \eqref{LZp} describes the $Z^\prime$ coupling to fermions.
We denote by $q_L^i$ and $\ell_L^i$  the  $\text{SU}(2)_L$ left-handed quark and lepton doublets, respectively,  with generation index $i = \{1,2, 3\}$. $u_R^i,\,d_R^i,\,\nu_R^i,\,e_R^i$ are  right-handed singlets.

Before the electroweak SSB  the $Z^\prime$ couplings to fermions are flavour conserving, hence   we can write 
 \be
{\cal L}^{Z^\prime}_{\rm int, fermions}=\sum_\psi {\cal L}_{\rm int}^{Z^\prime}
\ee
where the sum extends over all the fermions generically denoted by $\psi$ and
\be
{\cal L}_{\rm int}^{Z^\prime} = g_Z \, z_\psi \,\bar \psi \, \gamma^\mu \, \psi  \,Z_\mu^\prime \;. \label{Lint}
\ee
$g_Z$ is the $\text{U}(1)^\prime$ gauge coupling, and $z_\psi$  the $z$-hypercharge of the fermion $\psi$, i.e. the fermion quantum number  related to the new symmetry group. 
In  SM the fermions are chiral, hence it is useful to write  \eqref{Lint}  in terms of the left- and right-handed fermion fields $\psi_{L(R)}$:
\be
{\cal L}_{\rm int}^{Z^\prime} = \big[ (\Delta_L^\psi)^{ij} \,\bar \psi_L^i \, \gamma^\mu \, \psi_L^j  +(\Delta_R^\psi)^{ij} \,\bar \psi_R^i \, \gamma^\mu \, \psi_R^j \big] \, Z_\mu^\prime \;, \label{LintLR}
\ee
with
\be
(\Delta_{L,R}^\psi)^{ij} = g_Z \, z_{\psi_{L,R}} \, \delta^{ij} \;. \label{delta-couplings}
\ee
The $Z^\prime$ coupling to  the SM Higgs field $\varphi$ is described by ${\cal L}_{\varphi}^{Z^\prime}$. 
We write the covariant derivative acting on the SM Higgs field as ${\bar D}_\mu=D_\mu+i\,g_Z \,z_H\, Z^\prime_\mu$. $D_\mu$  contains only the SM gauge fields and $z_H$  is  the Higgs $z$-hypercharge.
 Therefore, we have 
\bea
({\bar D}_\mu \varphi)^\dagger ({\bar D}^\mu \varphi)&=&(D_\mu \varphi)^\dagger (D^\mu \varphi)+g_H \, \big( \varphi^\dagger \, i \, \overset{\leftrightarrow}{D}_\mu \,  \varphi \big)\, Z^{\prime \mu}\nn \\ &+&g_H^2 Z^\prime_\mu  Z^{\prime \mu } (\varphi^\dagger  \varphi) \,\,\, , \label{LHnew}
\eea
where $\varphi^\dagger \, i \, \overset{\leftrightarrow}{D}_\mu \,  \varphi =\varphi^\dagger \, ( i \,D_\mu \,  \varphi ) - ( i \, D_\mu \, \varphi^\dagger ) \,  \varphi $ and we have  defined
 \be
 g_H = g_Z \, z_H \;. \label{gh}
 \ee
 The last term in \eqref{LHnew} can be neglected in the present study since we are interested in  $d=6$ operators  arising at the EW scale when $Z^\prime$ is integrated out. The last term in \eqref{LHnew} produces a $d=8$ operator,  its coefficient would be  inversely proportional to $M_{Z^\prime}^4$.
 Therefore, in \eqref{LZp} we only include  
\be
\label{higgs_coupling}
{\cal L}_{\varphi}^{Z^\prime} = g_H \, \big( \varphi^\dagger \, i \, \overset{\leftrightarrow}{D}_\mu \,  \varphi \big)\, Z^{\prime \mu} \;.
\ee

\section{$\nu$SMEFT operators generated in the $\text{U}(1)^\prime$ extension of  SM }\label{smeft}
In the Warsaw basis the  operators are collected in classes according to their field content. 
The scalar field is denoted by $\varphi$, with $\tilde\varphi$ defined as
$\tilde\varphi^j=\epsilon_{jk}(\varphi^k)^*$  ($j,k$ are $\text{SU}(2)_L$ indices). 
The gauge field strengths are indicated  by $X$, $\tilde X$ being their duals. 
Fermions are  denoted by $\psi$. 
Among the various terms in ${\cal L}^{(6)}$ in Eq.~\eqref{LSMEFT}, we focus on ${\cal L}^{(6)}_{Z^\prime}$,  the set of operators generated at the EW scale when the SM group is extended including  $\text{U}(1)^\prime$ and the gauge boson $Z^\prime$ is integrated out. ${\cal L}^{(6)}_{Z^\prime}$ consists of the  terms\,\footnote{While in Eq.~\eqref{Ld} the Wilson coefficients are dimensionless, in \eqref{L6} it is   convenient to include the mass dimension in the definition of the coefficients. The operator ${\cal O}^{(6)}_{\nu \nu}$ is denoted by a superscript  to distinguish it from the $d=5$  Weinberg operator  ${\cal O}_{\nu \nu}$.}
\bea
\label{L6}
{\cal L}^{(6)}_{Z^\prime} &=& C_{\ell \ell} \, {\cal O}_{\ell \ell} + C_{qq}^{(1)} \, {\cal O}_{qq}^{(1)} \nn \\
&+& C_{ee} \, {\cal O}_{ee} + C_{uu} \, {\cal O}_{uu} + C_{dd} \, {\cal O}_{dd} + C^{(6)}_{\nu\nu} \, {\cal O}^{(6)}_{\nu \nu} \nn \\
&+& C_{\ell q}^{(1)} \, {\cal O}_{\ell q}^{(1)} + C_{ud}^{(1)} \, {\cal O}_{ud}^{(1)} + C_{eu} \, {\cal O}_{eu} + C_{ed} \, {\cal O}_{ed} \nn \\
&+& C_{\ell e} \, {\cal O}_{\ell e} + C_{\ell u} \, {\cal O}_{\ell u} + C_{\ell d} \, {\cal O}_{\ell d} + C_{qe} \, {\cal O}_{qe} +   \nn \\
&+&C_{ q u}^{(1)} \, {\cal O}_{q u}^{(1)} + C_{ q d}^{(1)} \, {\cal O}_{q d}^{(1)} + C_{\nu e} \, {\cal O}_{\nu e} + C_{\nu u} \, {\cal O}_{\nu u} \nn \\
&+& C_{\nu d} \, {\cal O}_{\nu d} + C_{\ell \nu} \, {\cal O}_{\ell \nu} + C_{q \nu} \, {\cal O}_{q \nu}  \\
&+& C_{\varphi \Box} \, {\cal O}_{\varphi \Box} + C_{\varphi D} \, {\cal O}_{\varphi D}  \nn \\
&+& C_{e \varphi} \,  {\cal O}_{e \varphi} + C_{u \varphi}  \, {\cal O}_{u \varphi} + C_{d \varphi}  \, {\cal O}_{d \varphi} + C_{\nu \varphi} \,  {\cal O}_{\nu \varphi} \nn \\
&+& C_{ \varphi \ell}^{(1)} \, {\cal O}_{\varphi \ell}^{(1)} + C_{ \varphi e} \, {\cal O}_{\varphi e} + C_{ \varphi q}^{(1)} \, {\cal O}_{\varphi q}^{(1)} \nn \\
&+& C_{ \varphi u} \, {\cal O}_{\varphi u} + C_{ \varphi d} \, {\cal O}_{\varphi d} + C_{\varphi \nu} \,  {\cal O}_{\varphi \nu} + \text{h.c.} \;. \nn
\eea
The various operators can be classified in the following classes  defined in \cite{Grzadkowski:2010es,Liao:2016qyd}:
\begin{itemize} 
\item four-fermion operators ${\cal O}_{\psi_1 \psi_2}$ (denoted as  ${\cal O}_{\psi\psi}$ if $\psi_1=\psi_2$) with structure $( {\bar L} L ) ( {\bar L} L )$:
\bea
 \big[ {\cal O}_{\ell \ell} \big]_{ijkp} &=& \big( {\bar \ell}_L^i \,  \gamma_\mu \,  \ell_L^j \big) \, \big( {\bar \ell}_L^k \,  \gamma^\mu \,  \ell_L^p \big)
 \nn \\
  \big[ {\cal O}_{qq}^{(1)} \big]_{ijkp} &=& \big( {\bar q}_L^i \,  \gamma_\mu \,  q_L^j \big) \, \big( {\bar q}_L^k \,  \gamma^\mu \,  q_L^p \big)
 \\
  \big[ {\cal O}_{\ell q}^{(1)} \big]_{ijkp} &=& \big( {\bar \ell}_L^i \,  \gamma_\mu \,  \ell_L^j \big) \, \big( {\bar q}_L^k \,  \gamma^\mu \,  q_L^p \big) \; ;
 \nn 
\eea
\item four-fermion operators ${\cal O}_{\psi_1 \psi_2}$ with structure $( {\bar R} R ) ( {\bar R} R )$:
\bea
 \big[ {\cal O}_{ee} \big]_{ijkp} &=& \big( {\bar e}_R^i \,  \gamma_\mu \,  e_R^j \big) \, \big( {\bar e}_R^k \,  \gamma^\mu \,  e_R ^p \big)
 \nn \\
 \big[ {\cal O}_{uu} \big]_{ijkp} &=& \big( {\bar u}_R^i \,  \gamma_\mu \,  u_R^j \big) \, \big( {\bar u}_R^k \,  \gamma^\mu \,  u_R ^p \big)
 \nn \\
 \big[ {\cal O}_{dd} \big]_{ijkp} &=& \big( {\bar d}_R^i \,  \gamma_\mu \,  d_R^j \big) \, \big( {\bar d}_R^k \,  \gamma^\mu \,  d_R ^p \big)
 \nn \\
 \big[ {\cal O}_{ud}^{(1)} \big]_{ijkp} &=& \big( {\bar u}_R^i  \, \gamma_\mu \,  u_R^j \big) \, \big( {\bar d}_R^k \,  \gamma^\mu \,  d_R ^p \big)
 \nn \\
 \big[ {\cal O}_{eu} \big]_{ijkp} &=& \big( {\bar e}_R^i \,  \gamma_\mu \,  e_R^j \big) \, \big( {\bar u}_R^k \,  \gamma^\mu \,  u_R ^p \big)
 \nn \\
 \big[ {\cal O}_{ed} \big]_{ijkp} &=& \big( {\bar e}_R^i \,  \gamma_\mu \,  e_R^j \big) \, \big( {\bar d}_R^k \,  \gamma^\mu  \, d_R ^p \big)
  \\
\big[ {\cal O}_{\nu \nu}^{(6)} \big]_{ijkp} &=& \big( {\bar \nu}_R^i \,  \gamma_\mu \,  \nu_R^j \big) \, \big( {\bar \nu}_R^k \,  \gamma^\mu \,  \nu_R ^p \big)
  \nn \\
 \big[ {\cal O}_{\nu e} \big]_{ijkp} &=& \big( {\bar \nu}_R^i \,  \gamma_\mu \, \nu_R^j \big) \, \big( {\bar e}_R^k \,  \gamma^\mu \,  e_R ^p \big)
 \nn \\
 \big[ {\cal O}_{\nu u} \big]_{ijkp} &=& \big( {\bar \nu}_R^i \,  \gamma_\mu \,  \nu_R^j \big) \, \big( {\bar u}_R^k \,  \gamma^\mu \, u_R ^p \big)
 \nn \\
 \big[ {\cal O}_{\nu d} \big]_{ijkp} &=& \big( {\bar \nu}_R^i \,  \gamma_\mu \,  \nu_R^j \big) \, \big( {\bar d}_R^k \,  \gamma^\mu \,  d_R ^p \big)  \; ; \nn
\eea
\item four-fermion operators ${\cal O}_{\psi_1 \psi_2}$ with structure $( {\bar L} L ) ( {\bar R} R )$:
\bea
 \big[ {\cal O}_{\ell e} \big]_{ijkp} &=& \big( {\bar \ell}_L^i \,  \gamma_\mu \,  \ell_L^j \big) \, \big( {\bar e}_R^k \,  \gamma^\mu \,  e_R ^p \big)
 \nn \\
 \big[ {\cal O}_{q e} \big]_{ijkp} &=& \big( {\bar q}_L^i \,  \gamma_\mu \,  q_L^j \big) \, \big( {\bar e}_R^k \,  \gamma^\mu \,  e_R ^p \big)
 \nn \\
 \big[ {\cal O}_{\ell u} \big]_{ijkp} &=& \big( {\bar \ell}_L^i \,  \gamma_\mu \,  \ell_L^j \big) \, \big( {\bar u}_R^k \,  \gamma^\mu \,  u_R ^p \big)\nn \\
\big[ {\cal O}_{\ell d} \big]_{ijkp} &=& \big( {\bar \ell}_L^i  \, \gamma_\mu \,  \ell_L^j \big) \, \big( {\bar d}_R^k \,  \gamma^\mu \,  d_R ^p )
\nn \\
  \big[ {\cal O}_{qu}^{(1)} \big]_{ijkp} &=& \big( {\bar q}_L^i  \, \gamma_\mu \,  q_L^j \big) \, \big( {\bar u}_R^k \,  \gamma^\mu  \, u_R ^p )
  \\
 \big[ {\cal O}_{qd}^{(1)} \big]_{ijkp} &=& \big( {\bar q}_L^i \,  \gamma_\mu \,  q_L^j \big) \, \big( {\bar d}_R^k \,  \gamma^\mu \,  d_R ^p )
 \nn 
\\
\big[ {\cal O}_{\ell \nu} \big]_{ijkp} &=& \big( {\bar \ell}_L^i \,  \gamma^\mu \,  \ell_L ^j \big) \,\big( {\bar \nu}_R^k \,  \gamma_\mu \,  \nu_R^p \big)  
 \nn \\
 \big[ {\cal O}_{q \nu} \big]_{ijkp} &=& \big( {\bar q}_L^i \,  \gamma^\mu \,  q_L ^j \big) \,\big( {\bar \nu}_R^k \,  \gamma_\mu \,  \nu_R^p \big)   \; ; \nn
 \eea
\item  operators ${\cal O}_{\varphi\partial}$ involving the Higgs field $\varphi$,  classified as    $\varphi^4 D^2$ in the Warsaw basis:
\bea
{\cal O}_{\varphi \square} &=& \big( \varphi^\dagger \, \varphi \big) \, \square \, \big( \varphi^\dagger \, \varphi \big) \nn \\
{\cal O}_{\varphi D} &=& \big( \varphi^\dagger \, D^\mu \, \varphi \big) \, \big( ( D_\mu \,  \varphi )^\dagger \, \varphi \big)  \;;
\eea
\item  operators ${\cal O}_{\psi\varphi}$ involving the Higgs field $\varphi$ and the fermion fields,  classified as    $\psi^2 \varphi^3$:
\bea
\big[ {\cal O}_{e \varphi} \big]_{ij} &=& \big( \varphi^\dagger \, \varphi \big) \, \big( {\bar \ell}_L^i \,  \varphi \, e_R^j \big) \nn \\
\big[ {\cal O}_{u \varphi} \big]_{ij} &=& \big( \varphi^\dagger \, \varphi \big) \, \big( {\bar q}_L^i \,  \tilde{\varphi} \,  u_R^j \big) \nn \\
\big[ {\cal O}_{d \varphi} \big]_{ij} &=& \big( \varphi^\dagger \,  \varphi \big) \, \big( {\bar q}_L^i \,  \varphi \,  d_R^j \big)  
\\
\big[{\cal O}_{\nu \varphi}\big]_{ij}& =& \big( \varphi^\dagger \, \varphi \big) \, \big( {\bar \ell}^i_L\, {\tilde  \varphi} \, \nu^j_R  \big)  \; ; \nn
\eea
\item  operators ${\cal O}_{\varphi \psi}$ comprising the Higgs field $\varphi$ and the fermion fields,  classified as    $\psi^2 \varphi^2 D$:
\bea
\big[ {\cal O}_{\varphi \ell}^{(1)} \big]_{ij} &=&
\big( \varphi^\dagger \, i \, \overset{\leftrightarrow}{D}_\mu \, \varphi \big) \, \big( {\bar \ell}_L^i \, \gamma^\mu \, \ell_L^j \big)
 \nn \\
 \big[ {\cal O}_{\varphi e} \big]_{ij} &=&
\big( \varphi^\dagger \, i \, \overset{\leftrightarrow}{D}_\mu \, \varphi \big) \, \big( {\bar e}_R^i \,  \gamma^\mu \,  e_R^j \big)\nn \\
\big[ {\cal O}_{\varphi q}^{(1)} \big]_{ij} &=&
\big( \varphi^\dagger \,  i \, \overset{\leftrightarrow}{D}_\mu \, \varphi \big) \, \big( {\bar q}_L^i \,  \gamma^\mu \,  q_L^j \big) \nn \\
\big[ {\cal O}_{\varphi u} \big]_{ij} &=&
\big( \varphi^\dagger \,  i \, \overset{\leftrightarrow}{D}_\mu \, \varphi \big) \, \big( {\bar u}_R^i \,  \gamma^\mu \,  u_R^j \big) \\
\big[ {\cal O}_{\varphi d} \big]_{ij} &=&
\big( \varphi^\dagger \,  i \, \overset{\leftrightarrow}{D}_\mu \, \varphi \big) \, \big( {\bar d}_R^i \,  \gamma^\mu  \, d_R^j \big)\nn 
\\
\big[{\cal O}_{ \varphi \nu }\big]_{ij} &=& \big( \varphi^\dagger \, i \, \overset{\leftrightarrow}{D}_\mu \, \varphi \big) \, \big( {\bar \nu}^i_R \, \gamma^\mu \, \nu^j_R \big) \;. \nn
\eea
\end{itemize}
 $i,j,k,p$ are generation indices. 

\section{Relations among the Wilson coefficients}\label{relations}

The coefficients of the operators  in Sec.\ref{smeft} can be expressed in terms of the couplings in Eq.~\eqref{LintLR} \cite{deBlas:2017xtg}. 
For  four-fermion operators they  read:
\bea
\big[ C_{\ell \ell} \big]_{ijkp} &=& - \frac{( \Delta_L^\ell )^{ij} \, ( \Delta_L^\ell )^{kp}}{2 \, M_{Z^\prime}^2}\label{cll} \\
\big[ C_{qq }^{(1)} \big]_{ijkp} &=& - \frac{( \Delta_L^q )^{ij} \, ( \Delta_L^q )^{kp}}{2 \, M_{Z^\prime}^2}\label{cqq} \\
\big[ C_{ee} \big]_{ijkp} &=& - \frac{( \Delta_R^e )^{ij} \, ( \Delta_R^e )^{kp}}{2 \, M_{Z^\prime}^2} \label{cee} \\
\big[ C_{uu} \big]_{ijkp} &=& - \frac{( \Delta_R^u )^{ij} \, ( \Delta_R^u )^{kp}}{2 \, M_{Z^\prime}^2} \label{cuu} \\
\big[ C_{dd} \big]_{ijkp} &=& - \frac{( \Delta_R^d )^{ij} \, ( \Delta_R^d )^{kp}}{2 \, M_{Z^\prime}^2} \label{cdd}
\eea
\bea
\big[ C^{(6)}_{\nu \nu} \big]_{ijkp} &=& - \frac{( \Delta_R^{\nu} )^{ij} \, ( \Delta_R^{\nu} )^{kp}}{2 \, M_{Z^\prime}^2} \label{cnn}
\eea
\bea
\big[ C_{\ell q }^{(1)} \big]_{ijkp} &=& - \frac{( \Delta_L^\ell )^{ij} \, ( \Delta_L^q )^{kp}}{M_{Z^\prime}^2}\label{clq} \\
\big[ C_{ud }^{(1)} \big]_{ijkp} &=& - \frac{( \Delta_R^u )^{ij} \, ( \Delta_R^d )^{kp}}{M_{Z^\prime}^2}\label{cud} \\
\big[ C_{eu } \big]_{ijkp} &=& - \frac{( \Delta_R^e )^{ij} \, ( \Delta_R^u )^{kp}}{M_{Z^\prime}^2}\label{ceu} \\
\big[ C_{ed } \big]_{ijkp} &=& - \frac{( \Delta_R^e )^{ij} \, ( \Delta_R^d )^{kp}}{M_{Z^\prime}^2}\label{ced} \\
\big[ C_{\ell e } \big]_{ijkp} &=& - \frac{( \Delta_L^\ell )^{ij} \, ( \Delta_R^e )^{kp}}{M_{Z^\prime}^2}\label{cle} \\
\big[ C_{\ell u } \big]_{ijkp} &=& - \frac{( \Delta_L^\ell )^{ij} \, ( \Delta_R^u )^{kp}}{M_{Z^\prime}^2}\label{clu} \\
\big[ C_{\ell d } \big]_{ijkp} &=& - \frac{( \Delta_L^\ell )^{ij} \, ( \Delta_R^d )^{kp}}{M_{Z^\prime}^2}\label{cld} \\
\big[ C_{q e } \big]_{ijkp} &=& - \frac{( \Delta_L^q )^{ij} \, ( \Delta_R^e )^{kp}}{M_{Z^\prime}^2}\label{cqe} \\
\big[ C_{qu }^{(1)} \big]_{ijkp} &=& - \frac{( \Delta_L^q )^{ij} \, ( \Delta_R^u )^{kp}}{M_{Z^\prime}^2}\label{cqu} \\
\big[ C_{qd }^{(1)} \big]_{ijkp} &=& - \frac{( \Delta_L^q )^{ij} \, ( \Delta_R^d )^{kp}}{M_{Z^\prime}^2}\label{cqd} 
\\
\big[ C_{\nu e } \big]_{ijkp} &=& - \frac{( \Delta_R^\nu )^{ij} \, ( \Delta_R^e )^{kp}}{M_{Z^\prime}^2}\label{cne} \\
\big[ C_{\nu u } \big]_{ijkp} &=& - \frac{( \Delta_R^\nu )^{ij} \, ( \Delta_R^u )^{kp}}{M_{Z^\prime}^2}\label{cnu} \\
\big[ C_{\nu d } \big]_{ijkp} &=& - \frac{( \Delta_R^\nu )^{ij} \, ( \Delta_R^d )^{kp}}{M_{Z^\prime}^2}\label{cnd} \\
\big[ C_{\ell \nu} \big]_{ijkp} &=& - \frac{( \Delta_L^\ell )^{ij} \, ( \Delta_R^\nu )^{kp}}{M_{Z^\prime}^2}\label{cln} \\
\big[ C_{q \nu} \big]_{ijkp} &=& - \frac{( \Delta_L^q )^{ij} \, ( \Delta_R^\nu )^{kp}}{M_{Z^\prime}^2}\label{cqn} \;.
\eea
%
The coefficients of the operators ${\cal O}_{\varphi \square}$ and ${\cal O}_{\varphi D}$ are given by
\bea
C_{\varphi \square} &=& - \frac{g_H^2}{2 \, M_{Z^\prime}^2}\label{cphiBox} \\
C_{\varphi D} &=& - \frac{2\,g_H^2}{M_{Z^\prime}^2}\;, \label{cphiD} 
\eea
so that
\bea
 C_{\varphi D} &=& 4 \, C_{\varphi \square} \,
\eea 
and  $C_{\varphi D} <0$.

The couplings to fermions enter  in the coefficients of ${\cal O}_{e \varphi}$, ${\cal O}_{u \varphi}$ and ${\cal O}_{d \varphi}$.
 However, when the UV completion consists only of the new $U(1)^\prime$ group, as considered in the present study, such coefficients vanish.
The coefficients of ${\cal O}_{\varphi \ell}^{(1)}$, ${\cal O}_{\varphi e}$, ${\cal O}_{\varphi q}^{(1)}$, ${\cal O}_{\varphi u}$ and ${\cal O}_{\varphi d} $ are given by
\bea
\big[ C_{\varphi \ell }^{(1)} \big]_{ij} &=& - \frac{( \Delta_L^\ell )^{ij} \, g_H}{M_{Z^\prime}^2}\label{cphil} \\
\big[ C_{\varphi e } \big]_{ij} &=& - \frac{( \Delta_R^e )^{ij} \, g_H}{M_{Z^\prime}^2}\label{cphie} \\
\big[ C_{\varphi q }^{(1)} \big]_{ij} &=& - \frac{( \Delta_L^q )^{ij} \, g_H}{M_{Z^\prime}^2}\label{cphiq} \\
\big[ C_{\varphi u } \big]_{ij} &=& - \frac{( \Delta_R^u )^{ij} \, g_H}{M_{Z^\prime}^2}\label{cphiu} \\
\big[ C_{\varphi d } \big]_{ij} &=& - \frac{( \Delta_R^d )^{ij} \, g_H}{M_{Z^\prime}^2}\label{cphid} \\
\big[ C_{\varphi \nu } \big]_{ij} &=& - \frac{( \Delta_R^\nu )^{ij} \, g_H}{M_{Z^\prime}^2} \;.\label{cphin}
\eea

For $N$  generations,  the coefficients in Eqs.~\eqref{cll}-\eqref{cqn} are  generally complex matrices in a $N^4$ dimensional space. 
However,
the coefficients in \eqref{cll}-\eqref{cnn} correspond to Hermitian operators, hence they are real and have $N^4$ components. 
In principle, the coefficients in Eqs.~ \eqref{cphil}-\eqref{cphin}  involve $2  N^2$ independent parameters. 
This parameter counting changes for the UV completion  obtained extending the SM gauge group with  the new $\text{U}(1)^\prime$.
We derive  relations among the coefficients before  SSB,  with unrotated  fermion fields and diagonal  $Z^\prime$ couplings to fermions. 
Moreover, in this case all  coefficients are real,  since they are expressed in terms of the (real) $z$-hypercharges and of $g_H$ which is real as from \eqref{gh}. 

Relations exist among the  remaining coefficients. We denote by $C_{\psi \psi}$ a generic coefficient among those in Eqs.~\eqref{cll}-\eqref{cnn}, and by $C_{\psi_1 \psi_2}$ a coefficient among  those in Eqs.~\eqref{clq}-\eqref{cqn}. 
The coefficients in Eqs.~\eqref{cphil}-\eqref{cphin} are generically denoted as $C_{\varphi \psi}$ (in all cases $\psi = \ell,\,q,\,\nu, \,e,\,u,\,d$).
We  have:
\bea
\big[ C_{\psi_1 \psi_2} \big]_{ijkp} &=& \pm 2 \, \sqrt{\big[C_{\psi_1 \psi_1}\big]_{ijij} \, \big[C_{\psi_2 \psi_2}\big]_{kpkp}} \qquad
\label{psi1psi2}\\
\big[C_{\psi \psi} \big]_{ijkp} &=& \frac{\big[C_{\varphi \psi} \big]_{ij} \, \big[C_{\varphi \psi} \big]_{kp}}{C_{\varphi D}} \;
\label{phi-psi-psi}\\
\big[ C_{\psi_1 \psi_2} \big]_{ijkp} &=& 2 \, \frac{\big[ C_{\varphi \psi_1} \big]_{ij} \, \big[ C_{\varphi \psi_2} \big]_{kp}}{C_{\varphi D}} \;.
\label{phi-psi1-psi2}
\eea
Considering Eq.~\eqref{delta-couplings},
 only the components $C_{ii kk}$ are nonvanishing among the coefficients  in \eqref{cll}-\eqref{cqn}. Moreover, the coefficients in \eqref{cll}-\eqref{cnn} are symmetric under the exchange ${ii} \leftrightarrow {kk}$, so they comprise only six independent components.
It is convenient to use the notation ${\underline i}={ii}\,,{\underline k}={kk}$. 
As for the coefficients $\left[C_{\varphi \psi}\right]_{ij}$ in Eqs.~\eqref{cphil}-\eqref{cphin}, they are nonvanishing only for $i=j$. 
We denote them as $\left[C_{\varphi \psi}\right]_{\underline i}$.\!\!\!
\footnote{To avoid confusion, when pedices refer to pairs of indices or to a single index we write ${\underline i}={\underline 1},{\underline 2}, {\underline 3}$ and $i=1,2,3$, respectively.}

Summarizing,   the following structures of coefficients are realized:
\begin{widetext}
\be
C_{\varphi \psi} =
\left(
\begin{array}{ccc}
\phantom{-}  \big[C_{\varphi \psi}\big]_{{\underline 1}} & \phantom{-} \big[C_{\varphi \psi}\big]_{{\underline 2}}\, & \phantom{-}  \big[C_{\varphi \psi}\big]_{{\underline 3}} \phantom{-} \\
\end{array} 
\right) \; \label{Cphipsimat-1}
\end{equation}
\be
C_{\psi \psi} =\frac{1}{C_{\varphi D}} \,
\left(
\begin{array}{ccc}
\phantom{-}\Big(\big[C_{\varphi \psi}\big]_{\underline 1}\Big)^2 & \phantom{-}  \big[C_{\varphi \psi}\big]_{\underline 1}\,\big[C_{\varphi \psi}\big]_{\underline 2} & \phantom{-} \big[C_{\varphi \psi}\big]_{\underline 1}\,\big[C_{\varphi \psi}\big]_{\underline 3} \phantom{-} \\
\vspace{-0.2cm} \\
\phantom{-} \big[C_{\varphi \psi}\big]_{\underline 2}\,\big[C_{\varphi \psi}\big]_{\underline 1} & \phantom{-}  \Big(\big[C_{\varphi \psi}\big]_{\underline 2}\Big)^2 & \phantom{-}  \big[C_{\varphi \psi}\big]_{\underline 2}\,\big[C_{\varphi \psi}\big]_{\underline 3}  \phantom{-}\\
\vspace{-0.2cm} \\
\phantom{-} \big[C_{\varphi \psi}\big]_{\underline 3}\,\big[C_{\varphi \psi}\big]_{\underline 1}& \phantom{-}  \big[C_{\varphi \psi}\big]_{\underline 3}\,\big[C_{\varphi \psi}\big]_{\underline 2} & \phantom{-}  \Big(\big[C_{\varphi \psi}\big]_{\underline 3}\Big)^2 \phantom{-} 
\end{array}
\right) \; \label{Cpsipsi-1}
\end{equation}
\be
C_{\psi_1 \psi_2} =\frac{2}{C_{\varphi D}} \,
\left(
\begin{array}{ccc}
\phantom{-}\big[C_{\varphi \psi_1}\big]_{\underline 1} \,\big[C_{\varphi \psi_2}\big]_{\underline 1}\quad & \phantom{-}  \big[C_{\varphi \psi_1}\big]_{\underline 1}\,\big[C_{\varphi \psi_2}\big]_{\underline 2} \quad& \phantom{-} \big[C_{\varphi \psi_1}\big]_{\underline 1}\,\big[C_{\varphi \psi_2}\big]_{\underline 3} \phantom{-} \\
\vspace{-0.2cm} \\
\phantom{-} \big[C_{\varphi \psi_1}\big]_{\underline 2}\,\big[C_{\varphi \psi_2}\big]_{\underline 1} \quad& \phantom{-}  \big[C_{\varphi \psi_1}\big]_{\underline 2}\, \big[C_{\varphi \psi_2}\big]_{\underline 2}\quad& \phantom{-}  \big[C_{\varphi \psi_1}\big]_{\underline 2}\,\big[C_{\varphi \psi_2}\big]_{\underline 3}   \phantom{-}\\
\vspace{-0.2cm} \\
\phantom{-} \big[C_{\varphi \psi_1}\big]_{\underline 3}\,\big[C_{\varphi \psi_2}\big]_{\underline 1} \quad& \phantom{-}  \big[C_{\varphi \psi_1}\big]_{\underline 3}\, \big[C_{\varphi \psi_2}\big]_{\underline 2}\quad& \phantom{-} \big[C_{\varphi \psi_1}\big]_{\underline 3}\,\big[C_{\varphi \psi_2}\big]_{\underline 3} \phantom{-} 
\end{array}
\right) \;. \label{Cpsi1psi2-1}
\end{equation}
\end{widetext}
The number of  independent coefficients in the dimension-six Lagrangian density \eqref{L6} is reduced to 19. They can be  
the 18 coefficients $\big[C_{\psi \psi}\big]_{{\underline i}\,{\underline i}}$ for ${\underline i}={\underline 1},{\underline 2},{\underline 3}$ and the six  $\psi = \ell,\,q,\,\nu , \, e,\,u,\,d$, and $C_{\varphi D}$; alternatively, they can be  the 18 coefficients $ \big[C_{\varphi \psi}\big]_{\underline i}$  and $C_{\varphi D}$.
In the next Section we describe the constraints for such coefficients  obtained requiring that the extended gauge group is free of gauge anomalies.

\section{Constraints from gauge anomaly cancellation}\label{anomalies}
 The issue of  gauge anomaly cancellation in presence of a new $\text{U}(1)^\prime$ symmetry  has been considered in many studies \cite{Carena:2004xs,Celis:2015ara,Ekstedt:2016wyi,Ismail:2016tod,Ellis:2017tkh,Allanach:2018vjg,Allanach:2019uuu,Aebischer:2019blw}.
In case of a new $Z^\prime$ gauge boson,  six gauge anomalies are generated.  They can be expressed  introducing the quantities $z_\psi^{(n)}$ defined in terms of the  sums  
\bea
z_\psi^{(n)} &=& \sum_{i = 1}^3 \, z_{\psi_i}^n \,\, , \label{znsum}
\eea
with $\psi_i$  a  fermion in the $i$ generation \cite{Carena:2004xs}.
The $[ \text{SU(3)}_C ]^2 \, \text{U(1)}^\prime$,  $[ \text{SU(2)}_L ]^2 \, \text{U(1)}^\prime$ and  $[ \text{U(1)}_Y ]^2 \, \text{U(1)}^\prime$
anomaly cancellation conditions
involve  the linear combinations of hypercharges in \eqref{znsum}, and read: 
\bea
A_{33z} &=& 2 \, z_q^{(1)} - z_u^{(1)} - z_d^{(1)} = 0 \; \label{A33z}
\\
A_{22z} &=& 3 \, z_q^{(1)} + z_\ell^{(1)} = 0 \; \label{A22z}
\\
A_{11z} &=& \frac{1}{6} \, z_q^{(1)} - \frac{4}{3} \, z_u^{(1)} - \frac{1}{3} \,z_d^{(1)} + \frac{1}{2} \, z_\ell^{(1)} - z_e^{(1)} = 0 . \qquad 
\label{A11z}
\eea
The triangular graph involving  two gravitons and $Z^\prime$ also produces a  relation linear in the $z$-hypercharges:
\be
A_{GGz} = 3 \, [ 2 \, z_q^{(1)} - z_u^{(1)} - z_d^{(1)} ] + 2 \, z_\ell^{(1)} - z_e^{(1)} - z_\nu^{(1)} = 0  \label{AGGza}
\ee
which can be simplified using Eq.~\eqref{A33z}:
\be
A_{GGz} = 2 \, z_\ell^{(1)} - z_e^{(1)} - z_\nu^{(1)} =0 \;. \label{AGGz}
\ee
The $\text{U(1)}_Y \, [ \text{U(1)}^\prime ]^2$ anomaly cancellation condition involves the quadratic sums in \eqref{znsum}:
\be
A_{1zz} = [ z^{(2)}_q -  2 \, z^{(2)}_u + z^{(2)}_d ]- [ z^{(2)}_\ell - z^{(2)}_e ] = 0  \;. \label{A1zz}
\ee
The  $[ \text{U(1)}^\prime ]^3$ anomaly cancellation condition involves the cubic sums in \eqref{znsum}:
\be
A_{zzz} = 3 \, [ 2 \, z^{(3)}_q -  z^{(3)}_u - z^{(3)}_d ] + [ 2 \, z^{(3)}_\ell -  z^{(3)}_\nu - z^{(3)}_e ] = 0 \;. \label{Azzz}
\ee

The previous equations provide constraints to  the coefficients in \eqref{L6}.
We  define 
\bea
{\tilde C}_{\varphi \psi}^{(n)} &=& \sum_{{\underline i}={\underline 1}}^{{\underline 3}} \, \Big( \big[ C_{\varphi \psi}\big]_{{\underline i}} \Big)^{n} \;,
\eea
denoting for simplicity
${\tilde C}_{\varphi \psi}^{(1)} = {\tilde C}_{\varphi \psi}$.
Using Eqs.~\eqref{cphil}-\eqref{cphin} we have
\be
z_{\psi_i} = - \frac{M_{Z^\prime}^2}{g_Z} \, \frac{1}{g_H} \, \big[ C_{\varphi \psi} \big]_{\underline i} \; ,
\ee
and Eq.~\eqref{znsum} becomes
\bea
z_\psi^{(n)}
&=& \left( - \frac{M_{Z^\prime}^2}{g_Z} \, \frac{1}{g_H} \right)^n \,  {\tilde C}_{\varphi \psi}^{(n)} \;. \label{zpsin}
\eea
With such definitions, the equations of the gauge anomaly cancellation conditions read:
\bea
A_{33z} &\to &2 \, {\tilde C}_{\varphi q} - {\tilde C}_{\varphi u} -  {\tilde C}_{\varphi d} = 0 \; \label{a33znew}
\\
A_{22z} &\to & 3 \, {\tilde C}_{\varphi q} + {\tilde C}_{\varphi \ell} = 0 \; \label{a22znew}
\\
A_{11z} &\to& {\tilde C}_{\varphi q} - 8 \, {\tilde C}_{\varphi u} - 2 \, {\tilde C}_{\varphi d} + 3 \, {\tilde C}_{\varphi \ell} - 6 \, {\tilde C}_{\varphi e} = 0 \;\,\,\,\,\,\,\,\,\,\,\label{a11znew} 
\\
A_{GGz} &\to& 2 \, {\tilde C}_{\varphi \ell} -  {\tilde C}_{\varphi e} -  {\tilde C}_{\varphi \nu} =0 \;.\label{aGGznew} 
\eea
They produce  the relations:
\bea
{\tilde C}_{\varphi q} &=& \frac{{\tilde C}_{\varphi u} + {\tilde C}_{\varphi d}}{2} \; \label{rel1phi}\\
{\tilde C}_{\varphi \ell} &=& - 3 \, {\tilde C}_{\varphi q} = - 3 \, \frac{{\tilde C}_{\varphi u}  + {\tilde C}_{\varphi d}}{2} \; \label{rel2phi}\\
{\tilde C}_{\varphi e} &=& - 2 \, {\tilde C}_{\varphi u} - {\tilde C}_{\varphi d} \; \label{rel3}
\\
{\tilde C}_{\varphi \nu} &=&  - {\tilde C}_{\varphi u}  -2 \, {\tilde C}_{\varphi d} \;. \label{rel4}
\eea
We also have
\bea
&&\hskip -.3cm
A_{1zz} \to {\tilde C}_{\varphi q}^{(2)} - 2 \, {\tilde C}_{\varphi u}^{(2)} + {\tilde C}_{\varphi d}^{(2)} - {\tilde C}_{\varphi \ell}^{(2)} +  {\tilde C}_{\varphi e}^{(2)} = 0 \quad \label{a1zznewphi}
\\
&&\hskip -.3cm
A_{zzz} \to 3 \, [ 2 \, {\tilde C}_{\varphi q}^{(3)} -  {\tilde C}_{\varphi u}^{(3)} - {\tilde C}_{\varphi d}^{(3)} ] \nn \\
& & \hspace{0.8cm} + [ 2 \, {\tilde C}_{\varphi  \ell}^{(3)} -  {\tilde C}_{\varphi  \nu}^{(3)} - {\tilde C}_{\varphi e}^{(3)} ] = 0 \;. \label{azzznewphi}
\eea
Examples on how the equations representing  the anomaly cancellation conditions (ACE) can be exploited are discussed below, considering models with  specific $z$-hypercharge assignments.

\section{Applications to models with specific $z$-hypercharge assignments}
The anomaly cancellation equations  involve 18 parameters: $\big[ C_{\varphi \psi}\big]_{{\underline i}}$ for  $\psi=q_i,\,\ell_i,\,u_i,\,d_i,\nu_i,\,\,e_i,$ and ${\underline i}={\underline 1},{\underline 2},{\underline 3}$.
Taking into account the constraints from  the 6 ACE, there are 12  independent parameters. 
With  other assumptions,  further constraints can be imposed,  as discussed below for selected cases.

\subsection{$Z^\prime$ only coupled  to the third generation, and $Z^\prime$  universally coupled to the three generations}
If $Z^\prime$ only couples  to one of the generations,  e.g. the third one,  we have $z_{\psi_3}=z_\psi$ and $z_{\psi_1}=z_{\psi_2}=0$. 
The number of parameters involved in the ACE is 6, denoted ${\bar C}_{\varphi \psi}$:
\bea
\big[ C_{\varphi \psi} \big]_{{\underline 3}}&=&{\bar C}_{\varphi \psi} \,\,\,, \hskip 0.5 cm \big[ C_{\varphi \psi} \big]_{{\underline 1}}=\big[ C_{\varphi \psi} \big]_{{\underline 2}}= 0 \,.
\eea
It follows that 
\bea
{\tilde C}_{\varphi \psi}^{(n)} &=&   \big(  {\bar C}_{\varphi \psi} \big)^{n} \;.
\eea

Before discussing the ACE, let us consider the scenario in which $Z^\prime$  universally couples to the three generations: 
$z_{\psi_1}=z_{\psi_2}=z_{\psi_3}=z_\psi$, as in models  where the $z$-hypercharge is a linear combination of the SM hypercharge $Y$ and of $B- n \, L$, with $B$ and $L$ the baryon and total lepton number and $n$ an integer number \cite{Carena:2004xs,Salvioni:2009jp,Accomando:2010fz}.\!\!\!
\footnote{Replacing $L$ with a family lepton number or a  combination of $L_e,\,L_\mu,\,L_\tau$ different from $L$  does not  belong to the generation independent category \cite{Salvioni:2009jp,Greljo:2021xmg}.}
The 6 parameters involved in the ACE are  denoted again by ${\bar C}_{\varphi\psi}$:
\bea
\big[ C_{\varphi \psi} \big]_{{\underline 1}}&=&\big[ C_{\varphi \psi} \big]_{{\underline 2}}=\big[ C_{\varphi \psi} \big]_{{\underline 3}}={\bar C}_{\varphi \psi}  \;,
 \eea
and the relation holds:
\bea
{\tilde C}_{\varphi \psi}^{(n)} &=&  3 \,  \big(  {\bar C}_{\varphi \psi} \big)^{n} \;.
\eea
The factor 3 factorises in the ACE, hence  the two cases are identical  from the viewpoint of solving  the equations  and can be discussed together.

\begin{figure}
\begin{center}
\includegraphics[width = 0.43\textwidth]{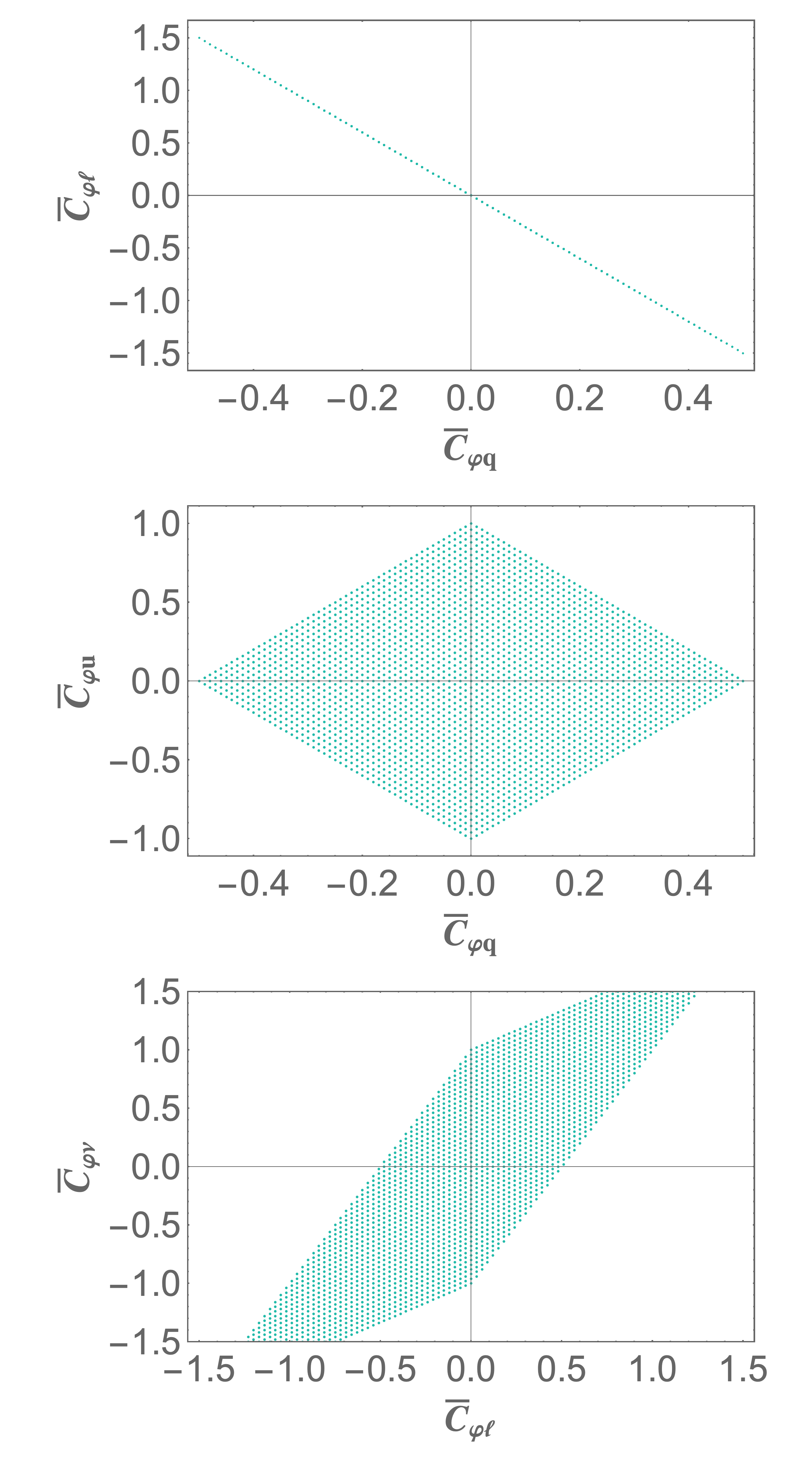}
\caption{\baselineskip 10pt  \small $Z^\prime$ only coupled  to the third fermion generation: Correlations among nonvanishing coefficients,  varying  ${\bar C}_{\varphi d}$ and ${\bar C}_{\varphi e}$ in the range $ [-1, 1]$.}\label{fig:cphiL3Gen}
\end{center}
\end{figure}

Since Eqs.~\eqref{a1zznewphi}-\eqref{azzznewphi} are automatically satisfied, they do not represent additional constraints, hence there are two independent coefficients.
One can express all  coefficients in terms of ${\bar C}_{\varphi d}$ and ${\bar C}_{ \varphi e}$:
\bea
{\bar C}_{\varphi q} &=& \frac{{\bar C}_{\varphi d}-{\bar C}_{\varphi e}}{4} \nn \\
{\bar C}_{\varphi u} &=& - \frac{{\bar C}_{\varphi d}+{\bar C}_{\varphi e}}{2} \nn \\
{\bar C}_{\varphi \ell} &=& - \frac{3 \, \big( {\bar C}_{\varphi d}-{\bar C}_{\varphi e} \big)}{4} \\
{\bar C}_{\varphi \nu} &=& \frac{- 3 \, {\bar C}_{\varphi d}+{\bar C}_{\varphi e}}{2} \nn \;.
\eea
Correlations among the four coefficients depending on the two independent ones are obtained, as shown in  Fig.~\ref{fig:cphiL3Gen} varying   ${\bar C}_{\varphi d}$ and ${\bar C}_{\varphi e}$.
\begin{figure}
\begin{center}
\includegraphics[width = 0.43\textwidth]{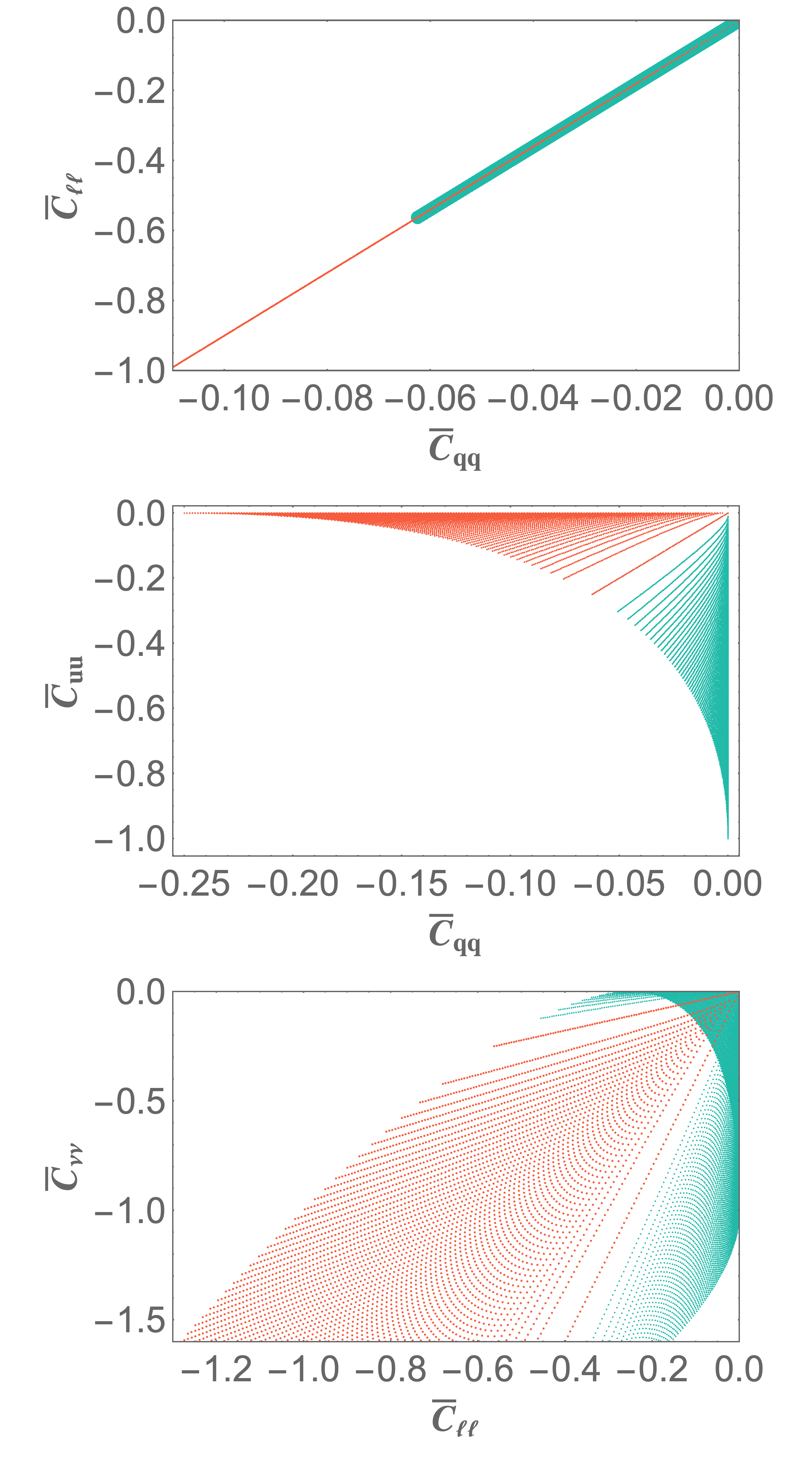}
\caption{\baselineskip 10pt  \small $Z^\prime$ only coupled  to the third generation: Correlation among nonvanishing coefficients, varying  ${\bar C}_{d d}$ and ${\bar C}_{e e}$ in the range $ [-1, 0]$. The green points refer to same-sign ${\bar C}_{d d}$ and ${\bar C}_{e e}$, the orange points to  the case of  opposite signs.}\label{fig:corr3genPsi}
\end{center}
\end{figure}

In this  specific scenario,  information can also be obtained on  $\left[C_{\psi \psi}\right]_{\underline{3} \, \underline{3}}$.
Indeed, Eqs.~\eqref{cll}-\eqref{cdd} imply 
\bea
z_{\psi_i} = \pm \left( 2 \, \frac{M_{Z^\prime}^2}{g_Z^2} \right)^{1/2} \, \Big( - \big[ C_{\psi \psi} \big]_{{\underline i}\,{\underline i}} \Big)^{1/2} \;. \label{zCpsirel}
\eea
As done for $C_{\varphi \psi}$ we define
%
$
\left[C_{\psi \psi}\right]_{\underline{3}\,\underline{3}}={\bar C_{\psi \psi}}.
$
The ACE can be used to relate the nonvanishing $z$-hypercharges:
\bea
z_{q_3}&=&\frac{1}{4} \, \big( z_{d_3}-z_{e_3} \big) \, \nn\\
z_{u_3}&=&-\frac{1}{2} \, \big( z_{d_3}+z_{e_3} \big)\,\nn\\
z_{\ell_3}&=&-{3 \over 4} \, \big( z_{d_3}-z_{e_3}\big)\,\\
z_{\nu_3}&=&{1\over 2} \,\big( z_{e_3}-3 \, z_{d_3} \big) \;. \nn
\eea
Two different cases can be analyzed, depending whether $z_{d_3}$ and $ z_{e_3}$ have same or opposite signs:
\begin{itemize} 
\item $z_{d_3}>0, \,\, z_{e_3}>0$ 
and $z_{d_3}<0,\,\, z_{e_3}<0$: we have
\bea
{\bar C_{q q}}&=& -\left({\sqrt{-{\bar C_{d d}}} - \sqrt{-{\bar C_{e e}}} \over 4}\right)^2 \,\nn\\
{\bar C_{u u}}&=& -\left({\sqrt{-{\bar C_{d d}}} + \sqrt{-{\bar C_{e e}}} \over 2}\right)^2\,\nn\\
{\bar C_{\ell \ell}}&=&-{9 \over 16} \, \left( \sqrt{-{\bar C_{d d}}}-\sqrt{-{\bar C_{e e}}}\right)^2\, \\
{\bar C_{\nu \nu}}&=&-\left( {\sqrt{-{\bar C_{e e}}}-3 \, \sqrt{-{\bar C_{d d}}} \over 2}\right)^2 \;,\nn
\eea
\item $z_{d_3}<0$,  $z_{e_3}>0$, and $z_{d_3}>0$, $z_{e_3}<0$: we have
\bea
{\bar C_{q q}}&=& -\left({\sqrt{-{\bar C_{d d}}} + \sqrt{-{\bar C_{e e}}} \over 4}\right)^2 \,\nn\\
{\bar C_{u u}}&=& -\left({\sqrt{-{\bar C_{d d}}} - \sqrt{-{\bar C_{e e}}} \over 2}\right)^2\,\nn\\
{\bar C_{\ell \ell}}&=&-{9 \over 16} \, \left( \sqrt{-{\bar C_{d d}}}+\sqrt{-{\bar C_{e e}}}\right)^2\, \\
{\bar C_{\nu \nu}}&=&-\left( {\sqrt{-{\bar C_{e e}}}+3\sqrt{-{\bar C_{d d}}} \over 2}\right)^2 \;. \nn 
\eea
\end{itemize}
Correlations among the four coefficients are obtained varying ${\bar C}_{d d}$ and ${\bar C}_{e e}$, as shown in Fig.~\ref{fig:corr3genPsi}. 

\subsection{$Z^\prime$ only coupled to left-handed fermions}
\begin{figure}[t]
\begin{center}
\includegraphics[width = 0.41\textwidth]{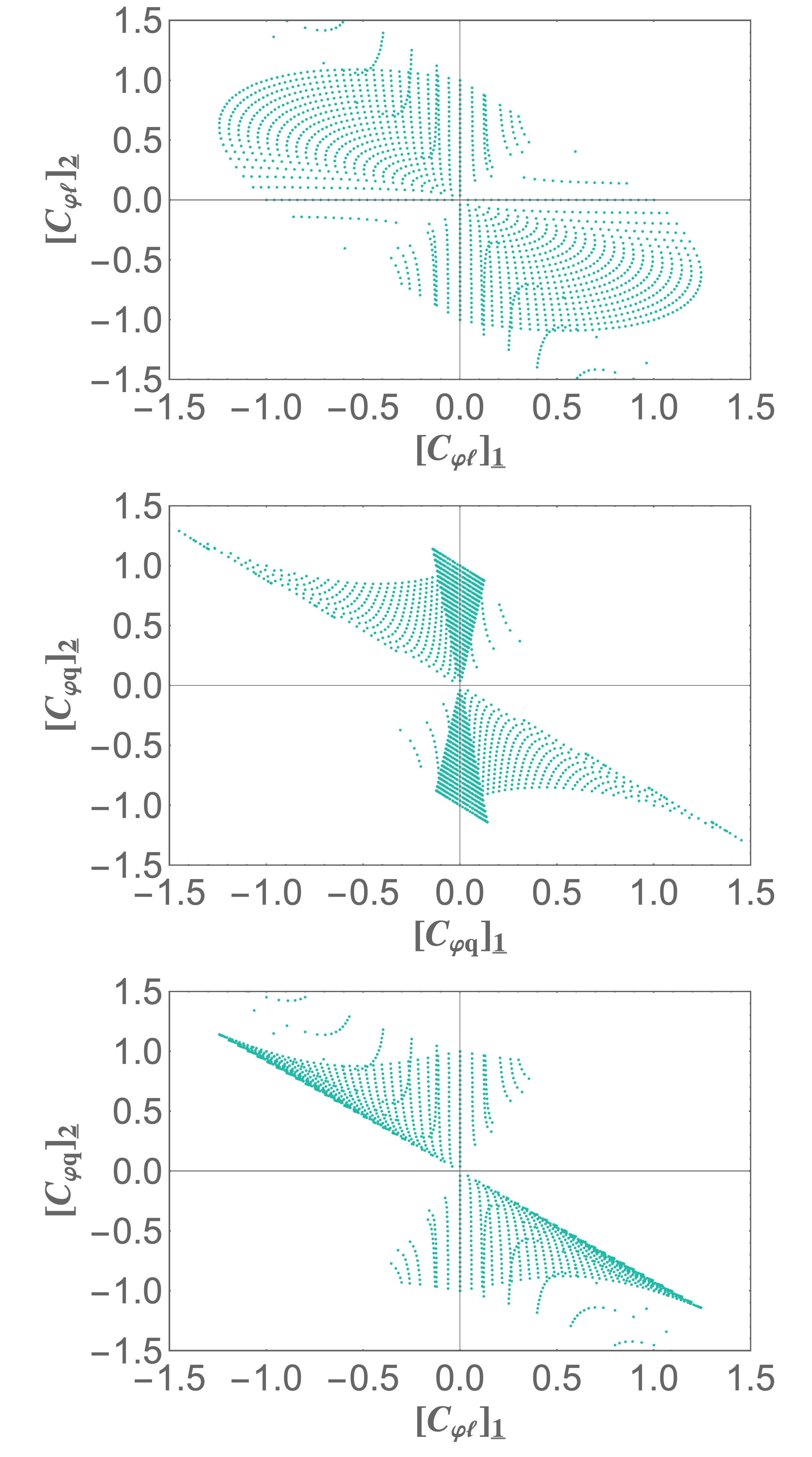}
\caption{\baselineskip 10pt  \small  $Z^\prime$ only coupled to left-handed fermions: Correlation among nonvanishing coefficients, varying  $[C_{\varphi \ell}]_{\underline 3}$ and $[C_{\varphi q}]_{\underline 3}$ in the range $[-1,1]$.}\label{fig:cphiL}
\end{center}
\end{figure}

The possibility that $Z^\prime$ only couples  to fermions of a given chirality has been considered,  e.g., in \cite{Buras:2012jb}.
If $Z^\prime$ only couples  to left-handed fermions the nonvanishing coefficients are $\big[ C_{\varphi \psi} \big]_{\underline i}$ for $\psi = \{ q_i,\,\ell_i \}$,  hence 6 parameters.
The number of constraints is reduced to 4 since Eqs.~\eqref{a22znew} and \eqref{a11znew} are redundant.
The linear equations \eqref{a33znew} and \eqref{aGGznew} provide the relations
\be
 {\tilde C}_{\varphi q} = {\tilde C}_{\varphi \ell} = 0 \;. \label{LH_fermions}
\ee
The quadratic and cubic ACE provide further relations, hence the number of independent coefficients is 2.
Varying   $[C_{\varphi \ell}]_{\underline 3}$ and $[C_{\varphi q}]_{\underline 3}$,  correlations are obtained among the remaining coefficients. They are shown in Fig.~\ref{fig:cphiL}.

\begin{figure}
\begin{center}
\includegraphics[width = 0.43\textwidth]{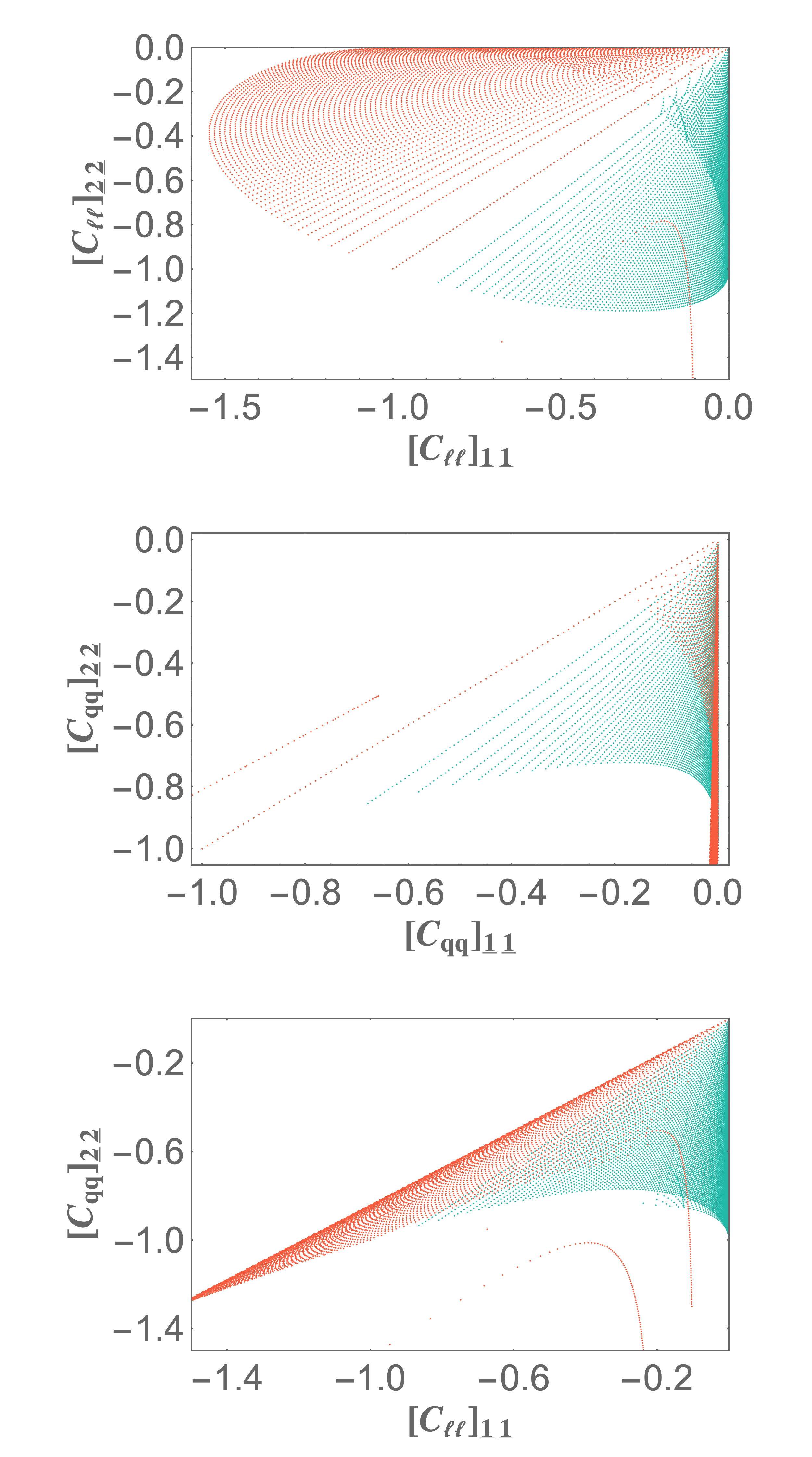}
\caption{\baselineskip 10pt  \small $Z^\prime$ coupled only to left-handed fermions: Correlations  among  coefficients,  varying $\left[ C_{\ell \ell}\right]_{\underline{3}\,\underline{3}}$ and $\left[ C_{q q}\right]_{\underline{3}\,\underline{3}}$ in the range $[-1,0]$. The color code  is the same as in   Fig.~\ref{fig:corr3genPsi}.}\label{fig:corrLeftpsi}
\end{center}
\end{figure}

Also in this case the ACE can be exploited to derive correlations among  $\left[C_{\psi \psi}\right]$,
choosing $\left[ C_{\ell \ell}\right]_{\underline{3}\,\underline{3}},\,\left[ C_{q q}\right]_{\underline{3}\,\underline{3}}$ as independent coefficients,  for same-sign or opposite-sign $z_{\ell_3}$ and $z_{q_3}$. 
The correlations  between the remaining coefficients    $\left[ C_{\ell \ell}\right]_{\underline{3}\,\underline{3}}$ and $\left[ C_{q q}\right]_{\underline{3}\,\underline{3}}$ are shown in  Fig.~\ref{fig:corrLeftpsi}.

\subsection{$Z^\prime$ only coupled to right-handed fermions}
If $Z^\prime$ only couples to right-handed fermions, the nonvanishing coefficients are $\big[ C_{\varphi \psi} \big]_{\underline i}$ for $\psi = \{ u_i,\,d_i, e_i, \nu_i \}$, hence 12 parameters.
The number of constraints is reduced to 5 since Eq.~\eqref{a22znew} is automatically satisfied.
Eqs.~\eqref{a33znew}, \eqref{a11znew} and \eqref{aGGznew} provide the relations
\be
 {\tilde C}_{\varphi u} = - {\tilde C}_{\varphi d} = - {\tilde C}_{\varphi e} = {\tilde C}_{\varphi \nu} \;. \label{RH_fermions}
\ee
The quadratic and cubic ACE give further relations, so that the number of independent coefficients is 7.

\subsection{Leptophobic $Z^\prime$}
If $Z^\prime$ only couples to quarks, the nonvanishing coefficients are $\big[ C_{\varphi \psi} \big]_{\underline i}$ for $\psi = \{ q_i,\,u_i,\,d_i \}$, therefore 9 parameters. 
The number of constraints is  5, since Eq.~\eqref{aGGznew} is automatically verified.
The other linear equations provide the relations
\be
 {\tilde C}_{\varphi q} = {\tilde C}_{\varphi u} = {\tilde C}_{\varphi d} = 0 \;, 
\ee
while the quadratic and cubic ACE read
\bea
&&{\tilde C}_{\varphi q}^{(2)} - 2 \, {\tilde C}_{\varphi u}^{(2)} + {\tilde C}_{\varphi d}^{(2)}  = 0 \;
\\
&&  2 \, {\tilde C}_{\varphi q}^{(3)} -  {\tilde C}_{\varphi u}^{(3)} - {\tilde C}_{\varphi d}^{(3)}  = 0 \;.
\eea
Consequently, there are 4   independent coefficients.

\subsection{Hadrophobic $Z^\prime$}
The situation is specular to the leptophobic $Z^\prime$. 
The expressions of the  ACE are
\bea
 &&{\tilde C}_{\varphi  \ell} = {\tilde C}_{\varphi  \nu} = {\tilde C}_{\varphi  e} = 0 \; \nn
\\
&&  - {\tilde C}_{\varphi  \ell}^{(2)} +  {\tilde C}_{\varphi  e}^{(2)} = 0 \; \label{hadroph}
\\
&&  2 \, {\tilde C}_{\varphi  \ell}^{(3)} -  {\tilde C}_{\varphi  \nu}^{(3)} - {\tilde C}_{\varphi  e}^{(3)}  = 0 \;.\nn 
\eea
The number of independent coefficients is 4.

For such models the experimental bounds are weaker than in previous cases, and allow a  relatively light $Z^\prime$. 
Moreover,  $Z^\prime$ can contribute to  lepton-flavour violating decays and to the lepton anomalous magnetic moments \cite{Heeck:2011wj,Harigaya:2013twa,Allanach:2015gkd,Altmannshofer:2016brv,Iguro:2020rby,Cheng:2021okr,Amaral:2021rzw,Kriewald:2022erk}, an issue of  great  interest  at present \cite{Muong-2:2021ojo,Muong-2:2023cdq}. 
Models gauging $L_a-L_b$ ($a,b$ being the lepton flavours)  belong to this class, namely models gauging   $L_\mu -L_\tau$ \cite{Altmannshofer:2014cfa,Crivellin:2015mga,Asai:2018ocx,Biswas:2019twf,Davighi:2020qqa,Buras:2021btx}.

 As an example of a hadrophobic model, we can also consider the  $Z^\prime$   only coupled to right-handed neutrinos, a scenario  belonging to the class of   {\it neutrinophilic}  NP models \cite{Huitu:2008gf,Basso:2008iv,Abdallah:2021npg}.
As for the ACE,  setting all z-hypercharges to $0$ but for right-handed neutrinos,  we have that   Eq.~\eqref{hadroph} is satisfied only if at least one of the three right-handed neutrinos is sterile under $\text{U}(1)^\prime$. 
Choosing $z_{\nu_3}=0$, the ACE imply $\left[C_{\varphi \nu} \right]_{\underline{3}} = 0$ and $\left[C_{\varphi \nu} \right]_{\underline{1}} = -\left[C_{\varphi \nu} \right]_{\underline{2}}$.

\subsection{ABCD model   \cite{Aebischer:2019blw}}
A  model with a heavy gauge boson  $Z^\prime$ with 
flavour nonuniversal quark and lepton couplings has been considered in \cite{Aebischer:2019blw}. 
The assignment of the $z$-hypercharge  to a generic  fermion  $\psi_i = \{ q_i,\,u_i,\,d_i,\,\ell_i,\,\nu_i,\,e_i \}$ ($i$ a generation index) is
\be
z_{\psi_i}= y_\psi +\epsilon_i \;.
\label{abcd}
\ee
$y_\psi$ denote the generation universal SM hypercharges,  $\epsilon_i$ are parameters  generation dependent,  but universal within a given generation. This construction produces   quark-lepton correlations. As shown in \cite{Aebischer:2019blw},  all  ACE are satisfied provided  
\be
\sum_{i=1}^{3} \epsilon_i=0 \;.
\ee
The assignment implies the  relation
\bea
{\tilde C}_{\varphi \psi} &=& 3 \, \left(-{g_Z^2 \, C_{\varphi D} \over 2 \, M_{Z^\prime}^2}\right)^{1/2} \, y_{\psi} \;.\label{abcd-phipsi}
\eea
For right-handed neutrinos  one has  ${\tilde C}_{\varphi \nu} = 0$ since $y_\nu= 0$.
Nontrivial relations among the SMEFT coefficients are predicted:
\be
- 6 \, \frac{{\tilde C}_{\varphi q} }{{\tilde C}_{\varphi e} } = - \frac{3}{2} \,\frac{{\tilde C}_{\varphi u} }{{\tilde C}_{\varphi e}} = 3 \, \frac{{\tilde C}_{\varphi d} }{{\tilde C}_{\varphi e}} = 2 \, \frac{{\tilde C}_{\varphi \ell} }{{\tilde C}_{\varphi e}} \;.
\ee

\section{Conclusions}
The possibility of gaining information on possible extensions of the SM, in a bottom-up approach, is largely based on the  SM effective field theory framework.  
It is important to obtain the widest  information  from the phenomenological analysis of the coefficients of the operators  in the effective field theory  Lagrangian. 
We have discussed the set of constraints and relations among the coefficients of the 
 $d=6$ operators if the SM extension includes a non-anomalous  $\text{U}(1)^\prime$. 
 In particular, we have investigated how the anomaly cancellation equations, involving the $z$-hypercharges, can be translated into constraints for the $\nu$SMEFT Wilson coefficients. 
 Such constraints become more stringent if particular features  are assumed for the  $Z^\prime$ couplings to fermions. 
 
 We have discussed  examples on how  the constraints can be exploited, and which correlations among the coefficients emerge.    Correlations among different coefficients imply relations among different physical processes, which can be searched and tested in experiment. Such processes could also involve neutrinos, which motivates our choice of considering the $\nu$SMEFT formulation. The correlations could also be included in global fit analyses using the data already available, or that will be collected in the near future. This provides us with  a way for accessing the long-sighted extension of the Standard Model.
 

\acknowledgements

We thank A.J. Buras and P. Stangl for  discussions.
This study has been  carried out within the INFN project (Iniziativa Specifica) SPIF.
The research has been partly funded by the European Union – Next Generation EU through the research Grant No. P2022Z4P4B “SOPHYA - Sustainable Optimised PHYsics Algorithms: fundamental physics to build an advanced society" under the program PRIN 2022 PNRR of the Italian Ministero dell’Universit\'a e Ricerca (MUR).

\bibliographystyle{apsrev4-1}
\bibliography{refSMEFT}
\end{document}